\pdfoutput=1 
\documentclass[aps,prl,twocolumn,superscriptaddress,amsfont,graphicx,nofootinbib,preprintnumbers]{revtex4}

\usepackage{color,graphicx,epsfig}
\usepackage{ifpdf}
\usepackage{amsmath}
\usepackage{bm}
\usepackage{color}
\usepackage[english]{babel}
\usepackage{graphicx}
\usepackage{amsfonts}
\usepackage{amssymb}
\usepackage{braket}
\usepackage{hyperref}
\usepackage{enumerate}
\usepackage{capt-of}
\usepackage{tikz}
\usepackage[caption=false]{subfig}

\usepackage{array, booktabs, makecell, multirow}

\definecolor{nicered}{rgb}{0.7,0.1,0.1}
\definecolor{nicegreen}{rgb}{0.1,0.5,0.1}
\hypersetup{colorlinks,citecolor=nicegreen,linkcolor=nicered,urlcolor=nicered}

\newcommand{\as}{\alpha_s}

\newcommand{\asb}{\alpha_{s,b}}

\newcommand{\ep}{\epsilon}

\newcommand{\be}{\begin{equation}}
\newcommand{\ee}{\end{equation}}
\newcommand{\bea}{\begin{eqnarray}}
\newcommand{\eea}{\end{eqnarray}}
\definecolor{Red}{rgb}{1.,0.,0.}
\definecolor{randomcolor}{rgb}{0.2,0.5,0.7}

\newcommand{\fin}{{\rm fin}}

\def\Tr{t}
\newcommand{\za}[1]{\resizebox{17pt}{!}{$\langle #1 \rangle$}}
\newcommand{\zb}[1]{\resizebox{15pt}{!}{$[#1]$}}

\DeclareMathOperator{\tr}{tr}

\DeclareMathAlphabet\mathbfcal{OMS}{cmsy}{b}{n}

\newcommand{\colorvector}[1]{\mathbfcal{#1}}

\def\OMIT#1{}

\definecolor{darkred}{rgb}{0.9,0,0}

\definecolor{darkgreen}{rgb}{0,100,0}

\definecolor{darkblue}{rgb}{0,0,0.9}

\allowdisplaybreaks[1]

\begin{document}
\def\OX{Rudolf Peierls Centre for Theoretical Physics, University of Oxford, Clarendon Laboratory, Parks
Road, Oxford OX1 3PU}
\def\TUM{Physik Department,
James-Franck-Straße 1, Technische Universit\"at M\"unchen,
D–85748 Garching, Germany}
\def\ORG{Exzellenzcluster ORIGINS, Boltzmannstr. 2, D-85748 Garching, Germany}
\def\MSU{Department of Physics and Astronomy, Michigan State University, East Lansing, Michigan 48824, USA}
\def\WDM{Wadham College, University of Oxford, Parks Road, Oxford OX1 3PN, UK}
\def\NEW{New College, University of Oxford, Holywell Street, Oxford OX1 3BN, UK}
\def\UR{Institut für Theoretische Physik, Universit\"at Regensburg, 93040 Regensburg, Germany}
\preprint{KA-TP-27-2023, MSUHEP-23-030, OUTP-23-12P, P3H-23-088, TUM-HEP-1481/23}

\title{Five-Parton Scattering in QCD at Two Loops}

\author{Bakul Agarwal}            
\email[Electronic address: ]{bakul.agarwal@kit.edu}
\affiliation{Institute for Theoretical Physics, Karlsruhe Institute of Technology (KIT), D-76131 Karlsruhe, Germany}

\author{Federico Buccioni}            
\email[Electronic address: ] {federico.buccioni@tum.de}
\affiliation{\TUM}

\author{Federica Devoto}            
\email[Electronic address: ] {federica.devoto@physics.ox.ac.uk}
\affiliation{\OX}

\author{Giulio~Gambuti}            
\email[Electronic address: ]{giulio.gambuti@physics.ox.ac.uk}
\affiliation{\OX}
\affiliation{\NEW}

\author{Andreas von Manteuffel}            
\email[Electronic address: ]{manteuffel@ur.de}
\affiliation{\UR}
\affiliation{\MSU}

\author{Lorenzo Tancredi}            
\email[Electronic address: ]{lorenzo.tancredi@tum.de}
\affiliation{\TUM}

\begin{abstract}
We compute all helicity amplitudes for the scattering of five
partons in two-loop QCD in all the relevant flavor configurations, retaining all contributing color structures. 
We employ tensor projection to obtain helicity amplitudes in 
the 't Hooft-Veltman scheme starting from a set of primitive amplitudes. 
Our analytic results are expressed in terms of massless pentagon functions, and are easy to
evaluate numerically. 
These amplitudes provide important input to investigations of soft-collinear factorization and to studies of the high-energy limit. 
\end{abstract}

\maketitle
\section{Introduction }

The calculation of scattering amplitudes for $n$ partons in perturbative Quantum Chromodynamics (QCD) has attracted much attention since the discovery of the ubiquitous role of Yang-Mills theories in the description of particle interactions.
These amplitudes constitute the building blocks for the calculation of cross sections for processes involving jets at hadron colliders, which have played a crucial role in providing direct experimental access to fundamental parameters of QCD such as the number of colors and the strong coupling constant.
At the LHC, the measurement of multi-jet cross sections at large transverse momenta constitutes a unique opportunity to explore QCD dynamics in extreme regimes~\cite{ATLAS:2011qvj,CMS:2013vbb,CMS:2014mna,ATLAS:2014qmg}.
Such high-precision analyses need to be matched by accurate theoretical predictions 
which as of today have been carried out 
to second order in perturbative QCD~\cite{Gehrmann-DeRidder:2019ibf,Czakon:2019tmo,Czakon:2021mjy,Chen:2022tpk,Alvarez:2023fhi}.
In addition, analytic calculations of $n$-parton scattering amplitudes provide important insights into the fundamental properties of Yang-Mills theories, such as their high-energy (Regge) limit (see $e.g.$~\cite{Gribov:2009zz}) or the universal structure of infrared divergences and factorization in soft and collinear limits (see $e.g.$ \cite{Agarwal:2021ais}).

Feynman diagram based calculations for multi-parton scattering amplitudes become challenging at higher orders in perturbation theory, 
and the relative simplicity of their results is often obscured by the complexity of the intermediate expressions.
More recently, new methods have been developed to exploit this simplicity and render higher order calculations manageable.
Standard techniques based on integration-by-parts identities (IBPs)~\cite{Tkachov:1981wb,Chetyrkin:1981qh,Laporta:2001dd} and differential equations~\cite{Kotikov:1990kg,Remiddi:1997ny,Gehrmann:1999as} have been augmented by
finite-field methods~\cite{vonManteuffel:2014ixa,Peraro:2019svx} and the use of canonical bases~\cite{Henn:2013pwa} to partly bypass heavy use of computer algebra and 
enable new ways to calculate loop integrals and amplitudes.
Moreover, better understanding of the mathematical properties of special functions~\cite{Remiddi:1999ew,Goncharov:1998kja,Goncharov:2001iea,Goncharov:2010jf} 
defined as iterated integrals~\cite{Chen:1977oja} has made it possible to devise efficient techniques 
for analytic and numerical evaluation of the ensuing integrals.

Thanks to these developments, a large number of previously unthinkable calculations have become possible, opening the way to entire new opportunities to test perturbative QCD. 
In particular, QCD form factors have now been computed to an astonishing four loops~\cite{Lee:2021uqq,Lee:2022nhh}, 
scattering amplitudes for four strongly interacting partons to three loops~\cite{Caola:2020dfu,Caola:2021izf,Caola:2021rqz},
and five-parton scattering up to two loops, largely in the leading-color approximation~\cite{Abreu:2018zmy,Abreu:2019odu,Badger:2013gxa,Badger:2015lda,Badger:2018enw,Badger:2019djh,Abreu:2021oya}.
For the latter, while all the ingredients have been available for some time, the corresponding sub-leading-color contributions have remained elusive given the algebraic complexity introduced by the non-planar diagrams.
More recently, full-color calculations have been completed for 
five-particle scattering processes involving fewer colored particles~\cite{Agarwal:2021vdh,Badger:2021ohm,Badger:2023mgf,Abreu:2023bdp}.

In this Letter, we remove the last remaining roadblock and address the calculation of complete two-loop corrections for the scattering of five partons in QCD. 
We consider all relevant partonic channels, $i.e.$\ the scattering of five gluons, of two quarks and three gluons, and of four quarks and one gluon, both for identical and different quark flavors.
Employing a combination of sophisticated computational methods, we derive analytic results expressed in terms of relatively simple rational functions and so-called pentagon functions for massless particles~\cite{Chicherin:2020oor}.
We provide all independent helicity amplitudes for the relevant partonic channels in
electronic format.

\allowdisplaybreaks

\section{Kinematics and color decomposition}
We consider the processes
\begin{align}\label{eq:process}
0 &\to g(p_1) + g(p_2) + g(p_3) + g(p_4) + g(p_5) , \nonumber\\
0 &\to \bar q (p_1) + q(p_2) + g(p_3) + g(p_4) + g(p_5) , \nonumber\\
0 &\to \bar q(p_1) + q(p_2) + \bar q'(p_3) + q'(p_4) + g(p_5) ,
\end{align}
where all momenta are outgoing and massless
\begin{equation}\label{eq:mom_cons}
p_1^\mu + p_2^\mu  + p_3^\mu + p_4^\mu + p_5^\mu = 0, \quad p_i^2 = 0,
\end{equation}
and $q$ and $q'$ correspond to strictly different flavors of quarks.
All other channels, including those involving four identical quarks, 
can be reconstructed from these by suitable permutations of the external momenta, 
as described below.

The five-point kinematics can be parameterized by five independent 
Mandelstam invariants $s_{ij} \equiv (p_i + p_j)^2$. We choose the cyclic set
\begin{align}\label{eq:mandelstams}
s_{12}, \; s_{23}, \; s_{34}, \; s_{45}, \; s_{51},
\end{align}
and identify the \textit{physical region} with
\begin{equation}
    s_{12}, s_{34}, s_{45} >0 ,\quad  s_{23}, s_{51} < 0
\end{equation}
which corresponds to the $12\to345$ scattering process. 

In order to describe all relevant helicity configurations 
we also employ the quantity 
\begin{equation}\label{eq:tr5def}
   \tr_5 = 4i\epsilon_{\mu\nu\rho\sigma} p_1^\mu p_2^\nu p_3^\rho p_4^\sigma \,,
\end{equation}
which has non-trivial transformation properties under parity, inherited from its definition through the Levi-Civita tensor as given in~\eqref{eq:tr5def}.

We employ \textit{dimensional regularization} to regulate 
both ultraviolet (UV) and infrared (IR) divergences. 
Specifically, we employ the 't Hooft-Veltman scheme (tHV)~\cite{tHooft:1972tcz},
which treats loop momenta in $d=4-2\ep$ dimensions, 
while retaining momenta and polarizations for external particles in four dimensions.

The bare amplitudes for each process in~\eqref{eq:process} can
be decomposed onto a set of color tensors $\mathcal{C}_c$ as
\begin{equation}\label{eq:physical_amplitude}
A = (4 \pi \asb)^{\frac{3}{2}} \colorvector{A} \cdot \colorvector{C} = (4 \pi \asb)^{\frac{3}{2}} \mathcal{A}_c \, \mathcal{C}_c \, ,
\end{equation}
where $\asb$ is the bare strong coupling.
In~\eqref{eq:physical_amplitude} $\mathcal{A}_c$ are color-ordered \emph{partial amplitudes}, $\mathcal C_c$ the corresponding elements of the color tensor basis and a summation over the index $c$ is implied. 
Let us introduce the shorthand notation
\begin{equation}
    \begin{split}  
        \Tr^{mn} &= \mathrm{Tr}[T^{a_m}T^{a_n}]\,, 
        \quad 
         \Tr^{mnl}_{i_1 i_2} = (T^{a_m} T^{a_n} T^{a_l})_{i_1 i_2}\, ,
        \\
        \Tr^{mn\dots k} &= \mathrm{Tr}[T^{a_m}T^{a_n}\dots T^{a_k}]-\mathrm{Tr}[T^{a_k}\dots T^{a_n}T^{a_m}]\,,
    \end{split}
\end{equation}
where the $SU(N_c)$ adjoint index $a_n$ is associated with the $n$-th external gluon, while the (anti-)fundamental index $i_n$ with the corresponding (anti-)quark state. 
The matrices $T^a_{ij}$ are generators of $SU(N_c)$ in the fundamental representation and they obey the normalization condition 
$\mathrm{Tr}[T^aT^b]
= \delta^{ab}/2$. With these definitions, the color basis for $ggggg$ reads
\begin{align}\label{eq:color_structures_ggggg}
    \{ \mathcal{C}_c & \}_{c=1,\dots,12} =  \{ 
    \Tr^{12345}, \;
    \Tr^{12354}, \;
    \Tr^{12435}, \;
    \Tr^{12453}, \;
    \Tr^{12534}, \;
    \notag\\
    &
    \Tr^{12543}, \;
    \Tr^{13245}, \;
    \Tr^{13254}, \;
    \Tr^{13425}, \;
    \Tr^{13524}, \;
    \Tr^{14235}, \;
    \Tr^{14325}
    \} ,
    \notag\\
    \{ \mathcal{C}_c & \}_{c=13,\dots,22} =  \{ 
    \Tr^{12}\Tr^{345},
    \Tr^{45}\Tr^{123},
    \Tr^{35}\Tr^{124},
    \Tr^{34}\Tr^{125},
    \notag\\
    & 
    \Tr^{13}\Tr^{245},
    \Tr^{25}\Tr^{134},
    \Tr^{24}\Tr^{135},
    \Tr^{14}\Tr^{235},
    \Tr^{23}\Tr^{145},
    \Tr^{15}\Tr^{234}
    \} \, ,
\end{align}
for $\bar{q}qggg$
\begin{align}
\label{eq:color_structures_qqggg}
\{ \mathcal{C}_c & \}_{c=1,\dots,6} = \{t^{345}_{i_1 i_2}, t^{453}_{i_1 i_2}, t^{534}_{i_1 i_2},
t^{354}_{i_1 i_2}, t^{543}_{i_1 i_2}, t^{435}_{i_1 i_2}\}, \notag \\
\{ \mathcal{C}_c & \}_{c=7,8,9} = \{T^{a_3}_{i_1 i_2} t^{45},\, T^{a_4}_{i_1 i_2} t^{53},\, T^{a_5}_{i_1 i_2} t^{34}\}, \notag \\
\mathcal{C}_{10,11} & = \delta_{i_1 i_2}\left(\mathrm{Tr}\left[T^{a_3}T^{a_4}T^{a_5}\right] \mp \mathrm{Tr}\left[T^{a_5}T^{a_4}T^{a_3}\right]\right) \,,
\end{align}
and finally for $ \bar q q \bar{q}' q'g $
\begin{equation}\label{eq:color_structures_qqqqg}
\begin{split}
&\mathcal{C}_1 = \delta_{i_2 i_3} T^{a_5}_{i_1 i_4}\,, \quad
\mathcal{C}_2 = \delta_{i_1 i_2} T^{a_5}_{i_3 i_4}\,,\\
&\mathcal{C}_3 = \delta_{i_3 i_4} T^{a_5}_{i_1 i_2}\,, \quad
\mathcal{C}_4 = \delta_{i_1 i_4} T^{a_5}_{i_3 i_2} \, . 
\end{split}
\end{equation}

\section{Helicity Amplitudes}

\begingroup
\renewcommand{\arraystretch}{2.3}
\begin{figure*}[th]
    \hspace{-1.5cm}
    \begin{minipage}[t]{0.30\textwidth}
        \centering
        \scalebox{0.90}{
            \begin{tabular}{cccc}
                $ ggggg$ & $\Phi(\bm \lambda)$ & $\mathcal{A}_c$ \\
                \hline
                \resizebox{30pt}{!}{$\bm{+++++}$}
                &
                $\displaystyle\frac{2s^2_{12}/3}{\za{12}\za{23}\za{34}\za{45}\za{51}}$
                & $1,13$ & 
                \\
                \resizebox{30pt}{!}{$\bm{-++++}$}
                &
                $\displaystyle\frac{  \zb{21}\za{12}^4 \za{13}^3}{\za{15}^2 \za{23}^5 \za{14}^2}$
                & $1,13$ 
                \\
                \resizebox{30pt}{!}{$\bm{--+++}$}
                &
                $\displaystyle\frac{4\za{12}^4}{\za{12}\za{23}\za{34}\za{45}\za{51}}$
                & $1,13$ 
                \\
                \resizebox{30pt}{!}{$\bm{-+-++}$}
                &
                $\displaystyle\frac{4\za{13}^4}{\za{12}\za{23}\za{34}\za{45}\za{51}}$
                & $1,13$ 
                \\
                \resizebox{30pt}{!}{$\bm{++++-}$}
                &
                $\displaystyle\frac{ \zb{51}\za{15}^4 \za{25}^3}{\za{54}^2 \za{12}^5 \za{53}^2}$
                & $13$\\
                \resizebox{30pt}{!}{$\bm{+++--}$}
                &
                $\displaystyle\frac{4\za{45}^4}{\za{12}\za{23}\za{34}\za{45}\za{51}}$ 
                & $13$
            \end{tabular}
        }
    \end{minipage}
    \begin{minipage}[t]{0.30\textwidth}
        \centering
        \vspace{-2.855cm}
        \scalebox{0.90}{
            \begin{tabular}{cccc}
                $ \bar{q} q ggg$ & $\Phi(\bm \lambda)$ & $\mathcal{A}_c$ \\
                \hline
                \resizebox{30pt}{!}{$\bm{+-++-}$}&
                $\displaystyle\frac{2\za{15}\za{52}^3}{\za{12}\za{23}\za{34}\za{45}\za{51}}$
                & $1,7,10,11$
                \\
                \resizebox{30pt}{!}{$\bm{+-+-+}$}&
                $\displaystyle\frac{2\za{14}\za{42}^3}{\za{12}\za{23}\za{34}\za{45}\za{51}}$
                & $1$
                \\
                \resizebox{30pt}{!}{$\bm{+--++} $}&
                $\displaystyle\frac{2\za{13}\za{32}^3}{\za{12}\za{23}\za{34}\za{45}\za{51}}$
                & $1$
                \\
                \resizebox{30pt}{!}{$\bm{+----}$}&
                $\displaystyle\frac{2\za{23}\zb{31}}{\zb{34}\zb{45}\zb{53}}$
                & $1,7,10,11$
                \\
                \resizebox{30pt}{!}{$\bm{+-+--}$}&
                $\displaystyle\frac{2\zb{23}\zb{31}^3}{\zb{12}\zb{23}\zb{34}\zb{45}\zb{51}}$
                & $7,10,11$
            \end{tabular}
        }
    \end{minipage} 
    \;
    \begin{minipage}[t]{0.30\textwidth}
        \centering
        \vspace{-2.855cm}
        \scalebox{0.90}{
             \begin{tabular}{ccccc}
                $\bar q q \bar q' q ' g$ & $\Phi(\bm \lambda)$ & $\mathcal{A}_c$ & $\Phi(\bm \lambda)$& $\mathcal{A}_c$ \\
                \hline
                \resizebox{30pt}{!}{$\bm{-+-+-}$}&
                $\displaystyle\frac{\zb{14} \zb{42}^2}{\zb{12}\zb{43}\zb{51}\zb{54}}$&
                $1$ &
                $\displaystyle\frac{\zb{42}^2}{\zb{12}\zb{53}\zb{54}}$ & $2$\\
                \resizebox{30pt}{!}{$\bm{-+-++}$}&
                $\displaystyle\frac{\za{41}\za{13}^2}{\za{15}\za{21}\za{34}\za{45}}$&
                $1$ &
                $\displaystyle\frac{\za{13}^2}{\za{21}\za{35}\za{45}}$ & $2$ \\
                \resizebox{30pt}{!}{$\bm{-++--}$}&
                $\displaystyle \frac{\zb{32}^2\zb{41}}{\zb{12}\zb{43}\zb{51}\zb{54}}$&
                $1$ &
                $\displaystyle \frac{\zb{32}^2}{\zb{21}\zb{53}\zb{54}}$ & $2$\\
                \resizebox{30pt}{!}{$\bm{-++-+}$}&
                $\displaystyle \frac{\za{14}^3}{\za{15}\za{21}\za{34}\za{45}}$&
                $1$ &
                $\displaystyle \frac{\za{14}^2}{\za{12}\za{35}\za{45}}$ & $2$
            \end{tabular}
        }
    \end{minipage}
    \captionof{table}{Definition of the minimal set of helicity and color configurations for the different processes. The list of helicity configuration is given in the left-most column of each table. The middle columns contain our definition of the spinor factors $\Phi_c({\boldsymbol\lambda})$ for the different color factors. The right columns give the list of partial amplitudes we computed analytically for the corresponding helicity configuration. Spinor factors are chosen to set the relative tree-level (when non-vanishing) to 1. Because of this, two spinor factor choices are needed for the $\bar q q \bar q ' q' g$ channel.}
    \label{tab:spins}
\end{figure*}
\twocolumngrid
\endgroup
We consider the scattering amplitudes for fixed helicity configurations of the external particles. We work in the spinor-helicity formalism and define the polarization vectors for 
gluons as 
\begin{equation}\label{eq:polvec_gluon}
\epsilon_{i,-}^\mu = \frac{[i+1|\gamma^\mu|i\rangle}{\sqrt{2} [i|i+1]}, \quad\quad \epsilon_{i,+}^\mu = \frac{[i|\gamma^\mu|i+1\rangle}{\sqrt{2} \langle i+1|i\rangle},
\end{equation}
and the spinors for (anti-)quarks as
\begin{equation}\label{eq:polvec_quark}
\overline{u}_{i,-} = \langle i| , \quad
\overline{u}_{i,+} = [i|   ,\quad
u_{i,-} = |i\rangle , \quad
u_{i,+} = |i]\, .
\end{equation}
Helicities are given in the all-outgoing convention.

We define spinor-stripped helicity amplitudes $\mathcal{H}_c(\boldsymbol\lambda)$ as
\begin{equation} \label{eq:spinor_stripped_ampls}
    \mathcal{A}_c(\boldsymbol{\lambda}) = \sqrt{2}\,
    \Phi_c({\boldsymbol\lambda})\,\mathcal{H}_c(\boldsymbol\lambda) \, ,
\end{equation}
where $\Phi_c({\boldsymbol\lambda})$ is a color and helicity dependent spinor factor that 
fully accounts for the little-group scaling of the corresponding amplitude. 
The scalar quantities $\mathcal{H}_c(\boldsymbol\lambda)$ can be further 
split into their parity even and odd parts 
\begin{equation}
     \mathcal{H}_c(\boldsymbol\lambda) = \mathcal{H}_c^E(\boldsymbol\lambda) + \tr_5 \mathcal{H}_c^O(\boldsymbol\lambda) 
\end{equation}
that are individually gauge invariant and can be computed independently.
We will only consider a minimal set of helicity and color configurations needed to reconstruct the whole amplitude via crossings of the external states. These are listed in tab.~\ref{tab:spins} along with the corresponding spinor factors.

The spinor-stripped helicity amplitudes contain both ultraviolet (UV) and infrared (IR) divergences, which  manifest as poles in the dimensional regulator $\ep$.
UV divergences can be removed by expressing the amplitudes in terms of the $\overline{\text{MS}}$ renormalized strong coupling $\as(\mu)$
\begin{equation}\label{eq:ren_coupling}
\asb \: \mu_0^{2\ep} \: S_\ep = \as(\mu) \:  \mu^{2\ep} \: Z\left[\as(\mu) \right],
\end{equation}
where $\mu_0$ and $\mu$ are the regularization and renormalization scales respectively, and 
$S_\ep = (4 \pi)^\ep e^{- \ep \gamma_E}$, with $\gamma_E \approx 0.5772$ the Euler constant.
Up to two loops, the renormalization factor $Z$ reads 
\begin{align}\label{eq:Zuv}
Z[\as]  &=  1
- \frac{\as}{4\pi}  \frac{ \beta_0 }{\epsilon } +\left( \frac{\as}{4\pi}\right)^2 \left( \frac{\beta_0^2}{\epsilon^2} - \frac{\beta_1 }{2 \epsilon} \right) \,,
\end{align}
with the $\beta$-function coefficients
\begin{equation}
    \begin{split}
        \beta_0 &= \frac{11}{3} N_c - \frac{2}{3} N_f \, , \\
        \beta_1 &= \frac{34}{3} N_c^2-\frac{10}{3} N_c N_f -\frac{N_c^2-1}{N_c} N_f \,,
    \end{split}
\end{equation}
where $N_c$ is the number of colors and $N_f$ the number of light fermions.
The renormalized helicity amplitudes can  be expanded as 
a perturbative series in the renormalized strong coupling
\begin{align}\label{eq:divergent_espansion}
\colorvector{A}(\bm{\lambda}) &= \sum_{\ell=0}^2 \left( \frac{\as}{4\pi}\right)^\ell \colorvector{A}_{\bm{\lambda},\text{ren}}^{(\ell)} + \mathcal{O}(\as^3),
\end{align}
where $\colorvector{A}_{\bm{\lambda},\text{ren}}^{(\ell)}$ is the renormalized $\ell$-loop contribution.
These still contain 
IR singularities, which can be subtracted by defining their finite remainders
at each loop order as
\begin{align}\label{eq:IR_subtraction}
        \colorvector{A}_{\bm{\lambda},\:\text{fin}}^{(0)} &= \colorvector{A}_{\bm{\lambda}}^{(0)} \; ,  \quad \colorvector{A}_{\bm{\lambda},\:\text{fin}}^{(1)} = \colorvector{A}_{\bm{\lambda},\: \text{ren}}^{(1)} - \mathbfcal{I}_1 \; \colorvector{A}_{\bm{\lambda},\: \text{ren}}^{(0)} \; ,   \nonumber\\ 
        \colorvector{A}_{\bm{\lambda},\:\text{fin}}^{(2)} &= \colorvector{A}_{\bm{\lambda},\: \text{ren}}^{(2)} - \mathbfcal{I}_2\; \colorvector{A}_{\bm{\lambda},\: \text{ren}}^{(0)} - \mathbfcal{I}_1\; \colorvector{A}_{\bm{\lambda},\: \text{ren}}^{(1)} \; . 
\end{align}
The color operators $\mathbfcal{I}_1$ and $\mathbfcal{I}_2$ were first defined in ref.~\cite{Catani:1998bh} and then in~\cite{Becher:2009cu,Becher:2009qa}.
For $\mathbfcal{I}_{1,2}$  we employ the definition from ref.~\cite{Becher:2009qa}.

\section{Details of the calculation}
To compute the helicity amplitudes up to two loops, we proceed as follows. 
We generate all Feynman diagrams for~\eqref{eq:process} using \texttt{QGRAF}~\cite{Nogueira:1991ex}. 
At two loops there are a total of 28020, 9136, and 2129 diagrams for 
the three channels respectively. 
However, not all of them contribute to the independent color structures described in eqs.~\eqref{eq:color_structures_ggggg},~\eqref{eq:color_structures_qqggg} and~\eqref{eq:color_structures_qqqqg}.  
The relevant diagrams can be selected by repeated application of the color identities
\begin{align}
        \mathrm{Tr} \,T^a &= 0, \quad T^a_{ij}T^a_{kh} = \frac{1}{2}\left(\delta_{ih}\delta_{kj} - \frac{1}{N_c}\delta_{ij}\delta_{kh} \right),\nonumber\\
        f^{abc} &= -2i\mathrm{Tr}(T^a[T^b,T^c]),
\end{align}
which reduce their color structure to a linear combination of the basis elements. 
From here, one can easily read off the contribution of the corresponding diagram 
to each element of the vector $\colorvector{A}$. 
We then compute the contribution of each diagram to the helicity amplitudes using 
projectors in the tHV scheme as in refs.~\cite{Peraro:2019cjj,Peraro:2020sfm}.
For each process we define helicity projectors $P_{\bm \lambda,c}$ as
\begin{equation}
    P_{\bm \lambda,c} \circ \mathcal{A}_c = \mathcal{H}_c(\bm \lambda)
\end{equation}
where the operation $``\circ"$ stands for summation over polarisations. 
The $P_{\bm \lambda,c}$ can be identified as follows. 
Each partial amplitude can be decomposed as
\begin{equation}
    \mathcal{A}_c(\bm \lambda) = \sum_{i=1}^{N} \mathcal{F}_{i,c} \: T_i(\bm \lambda)
\end{equation}
where the $T_i$ are a set of tensor structures containing all polarization vectors, and the $F_{i,c}$ are scalar \textit{form factors} for the different color structures. 
While the form of the decomposition is loop independent, the form factors themselves 
have a perturbative expansion and, at a given perturbative order, 
are linear combinations of scalar Feynman integrals. 
In our scheme, the number of tensor structures $N$ always equals the number of helicity amplitudes~\cite{Peraro:2019cjj,Peraro:2020sfm}. 
These are 32, 16, and 8 for the processes listed in~\eqref{eq:process}, respectively.
In practice this means that after numbering the helicity configurations ${\bm \lambda_1},{\bm \lambda_2},\dots$ for each process, we can change to a new basis of tensors 
$\overline{T}_{i,c}(\bm \lambda)$, which satisfy
\begin{equation}
     \overline{T}_{i,c}(\bm \lambda = \bm \lambda_j) = \delta_{ij} \Phi_c(\boldsymbol{\lambda}_j) \, .
\end{equation}
This allows us to define the helicity projectors $P_{\bm \lambda,c}$ by requiring
\begin{equation}
     P_{\bm \lambda,c} \circ \overline{T}_{i,c}(\bm \lambda') = \delta_{\bm\lambda,\bm\lambda'} \, .
\end{equation}
The Lorentz and color algebra required to
isolate the partial amplitudes and to apply the 
projectors are performed with \texttt{FORM}~\cite{Vermaseren:2000nd,Ruijl:2017dtg}.

As a result, the spinor-stripped helicity amplitudes 
$\mathcal{H}(\bm \lambda)$ in~\eqref{eq:spinor_stripped_ampls} are  expressed as linear combination of 
scalar Feynman integrals. All $L$-loop integrals required for the evaluation of 
these amplitudes are of the usual form
\begin{equation}\label{eq:integrals}
\mathcal{I}^{\text{fam}}_{n_1,...,n_N} = \mu_0^{2L\epsilon} e^{L \epsilon \gamma_E}  \int \prod_{i=1}^L \left( \frac{\mathrm{d}^d k_i}{i \pi^{\frac{d}{2}}} \right) \frac{1}{D_1^{n_1} \dots D_N^{n_N}}\,,
\end{equation}
where the loop momenta are labeled by $k_i$.  
The label ``fam" represents one of the two integral families $\{A,B\}$ 
as well as their crossed versions, which differ by a permutation of the external particles. 
Every integral family specifies the list of inverse propagators 
$D_e = q_e^2 + i\varepsilon$ in eq.~\eqref{eq:integrals}, where $\varepsilon$ implements the Feynman prescription.
Definitions for these integral families can be found in the supplemental material.
Collectively, across the various primitive amplitudes, we are left with 
$ \sim \mathcal{O}(10^6)$ different scalar integrals. 

It is well known that Feynman integrals satisfy many linear relations which can be 
obtained via symmetry relations and integration-by-parts identities (IBPs).
These can be used to express all Feynman integrals for a given family in terms of a set of basis integrals, referred to as \textit{master integrals}.
For the processes considered in this Letter, the set of master integrals was computed in refs.~\cite{Papadopoulos:2015jft,Gehrmann:2018yef,Chicherin:2018mue} via the method of differential equations, 
and expressed as a Laurent series in the dimensional regulator $\epsilon$. 
We use the uniform representation from ref.~\cite{Chicherin:2020oor} in terms of massless \textit{pentagon functions}.

In practice, we first derive shift relations and sector symmetries using \texttt{Reduze\;2}~\cite{Studerus:2009ye,vonManteuffel:2012np}. This allows us to reduce the number of integrals appearing in the unreduced amplitudes by two orders of magnitude.
We then perform IBP reduction using the public code \texttt{Kira}~\cite{Maierhofer:2017gsa,Klappert:2020nbg}
and, 
specifically for the non-planar topologies, the code \texttt{Finred}, an in-house implementation of Laporta's algorithm, employing finite field techniques~\cite{vonManteuffel:2014ixa,vonManteuffel:2016xki,Peraro:2016wsq,Peraro:2019svx}, syzygy algorithms~\cite{Gluza:2010ws,Schabinger:2011dz,Ita:2015tya,Larsen:2015ped,Boehm:2017wjc,Agarwal:2020dye}, and denominator guessing~\cite{Abreu:2018zmy,Heller:2021qkz}.

Although the number of integrals to reduce is substantially smaller compared to other cutting edge QCD calculations \cite{Caola:2020xup,Caola:2021izf,Caola:2021rqz,vonManteuffel:2020vjv,Lee:2022nhh,Gehrmann:2023jyv}, 
the multi-scale kinematics of five-point amplitudes is responsible for a large swell in intermediate expressions. 
In fact in the course of this calculation one encounters individual IBP identities 
with disk-sizes of up to $3$GB~\cite{Boehm:2020ijp,Agarwal:2021grm,Agarwal:2021vdh}.
A possible way around this consists in avoiding the reconstruction of IBP identities from finite-field samples for individual integrals and attempting, instead, to directly obtain the analytic expressions for the IR subtracted finite remainders of the various helicity amplitudes. 
As the finite remainders are expected to be simpler, this strategy has been very successful in many state-of-the-art five-point calculations.

In this work we follow a different approach~\cite{Agarwal:2021grm,Agarwal:2021vdh}, 
based on the observation that partial fraction decomposition 
can reduce expression size for single IBPs of various orders of magnitude.
We perform IBP reduction for a minimal subset of integrals and decompose the resulting integral coefficients into multivariate partial fractions.
Reduction identities for all integrals contributing to the amplitudes are 
then obtained using crossings of external invariants.
To perform the partial-fraction decomposition, we utilize \texttt{MultivariateApart}~\cite{Heller:2021qkz} augmented by \texttt{Singular}~\cite{DGPS}.
This step requires first defining an appropriate ordering among the denominator factors
that appear in the individual color factors of each helicity amplitudes. 
As expected, the complexity of the final expressions depends greatly on this choice 
of monomial ordering. 
We find it beneficial, in particular, to choose an ordering that avoids as much as possible 
the appearance of spurious singularities in the denominators,
for more details see refs.~\cite{Agarwal:2021grm,Agarwal:2021vdh}.
Once the simplified reduction identities are inserted into the amplitude, we perform an additional partial-fraction decomposition to simplify the resulting expressions further; in addition to the IBP reduction, this proves to be the step with the highest computational cost in our framework.
Once the bare amplitudes have been computed with a consistent choice of monomial ordering, UV renormalization and IR subtraction can be performed to yield relatively compact expressions for the finite remainders. 

A further level of simplification can be achieved by expressing the 
finite remainders in terms of a minimal set of independent rational functions.
Following refs.~\cite{Agarwal:2021grm,Agarwal:2021vdh}, we start from a partial-fractioned form of the rational functions based on a fixed monomial ordering. 
We exploit the uniqueness of this representation in order to search for linear relations among the rational functions, which we solve using row reductions.
As a measure of complexity we use the byte size of the rational functions and we 
reduce the system of linear relations to express more ``complicated" 
functions in terms of simpler ones, see ref.~\cite{Agarwal:2021vdh} for more details.
This final step allows us to obtain an improvement of around one order of magnitude in the disk-size of independent color-ordered primitive helicity amplitudes.

\section{Final Results and Conclusions}
The color and helicity configurations listed in eqs.~\eqref{eq:color_structures_ggggg},~\eqref{eq:color_structures_qqggg}, \eqref{eq:color_structures_qqqqg} and table~\ref{tab:spins} for the processes in eq.~\eqref{eq:process} are sufficient to reconstruct all possible partonic channels. 
In particular, using parity and charge conjugation, as well as permutations of the external particles,
one can obtain all amplitudes for the processes
\begin{equation} \label{eq:channels_all}
    \begin{split}
    &gg \to ggg, \quad
    q\bar q \to ggg, \quad
    q g \to qgg, \quad
    gg \to q \bar q g, \\
    &q\bar q\to \bar q' q' g, \;\;    
    q\bar q' \to  \bar q'  q g, \;\; 
    \bar q' \bar q \to \bar q' \bar q g, \;\;
    g \bar q \to  \bar q' q' \bar q   .
\end{split}
\end{equation}
The same-flavor quark amplitudes 
\begin{equation} \label{eq:channels_sameflavor}
    \begin{split}
        &
        q\bar q \to \bar q q g, \quad    
        q q \to q q g, \quad 
        qg \to qq\bar q
    \end{split}
\end{equation}
can then be obtained as linear combinations of appropriate components of the different-flavor quark amplitudes.
More specifically, in the notation of eq.~\eqref{eq:spinor_stripped_ampls}
we can write
\begin{equation}
    \begin{split}
            \colorvector{A}_{q\bar q \to q \bar q g}(\bm \lambda) =  \colorvector{A}_{q\bar q' \to q \bar q' g}(\bm \lambda) -  \colorvector{A}_{q\bar q \to q' \bar q' g}(\bm \lambda),\\
            \colorvector{A}_{qq \to q q g}(\bm \lambda) =  \colorvector{A}_{qq' \to q q' g}(\bm \lambda) -  \colorvector{A}_{qq' \to q' q g}(\bm \lambda),\\
            \colorvector{A}_{qg \to qq \bar q }(\bm \lambda) =  \colorvector{A}_{qg \to q q' \bar q' }(\bm \lambda) -  \colorvector{A}_{qg \to q' q \bar q' }(\bm \lambda),\\
    \end{split}
\end{equation}
where all channels on the rhs of these identities are either in eq.~\eqref{eq:channels_all} or can be obtained from them by crossings which do not require analytic continuation.

Crossings of rational functions and spinor factors amounts to a simple renumbering 
of momenta and their helicities (including flipping of helicities from incoming
to outgoing states).
Crossing of the transcendental functions, on the other hand, may require a dedicated analytic continuation.

We note that the analytic continuation of each pentagon function
individually is non-trivial but also not necessary.
In fact, the information needed to perform the required continuation 
is available
implicitly in the results provided by ref.~\cite{Chicherin:2020oor}, where all master integrals are evaluated for all $120$ permutations of the external invariants in terms of a minimal set of pentagon functions.
In practice, we took every uncrossed master integral and applied the required crossing on its analytic expression by formally crossing all appearing pentagon functions. 
We then equated these formal expressions to the crossed master integrals available in ref.~\cite{Chicherin:2020oor}, which are written in terms of uncrossed pentagon functions.
Repeating this for all master integrals for a given crossing, we obtained a linear system of equations, which we solved using \texttt{FiniteFlow}~\cite{Peraro:2019svx} to express the crossed pentagon functions in terms of the uncrossed ones.
Typically, the system is under-determined and some crossed pentagon functions remain unsolved for.
Nevertheless, since the results in ref.~\cite{Chicherin:2020oor} are, by construction, sufficient to represent any crossing of the amplitudes, all remaining crossed pentagon functions must cancel upon inserting these relations in the amplitude. 
We verified this cancellation explicitly for each crossing required to obtain the helicity and color ordered amplitudes for all partonic
sub-channels. 
This provided a strong check of the consistency of our procedure.

We performed numerous checks on our results.
First, we observed full cancellation of UV and IR poles of the bare amplitudes after UV renormalization and IR subtraction.
For the five-gluon channel, we also verified the one- and two-loop $U(1)$ decoupling identities as well as the generalized color-trace identities described in~\cite{Edison:2011ta}.
For the same channel, we also computed a redundant set of single-trace partial amplitudes; the crossing relations among them allowed us to verify the consistency of our calculation at the level of finite remainders.   
We also compared our tree-level and one-loop results against existing analytic calculations~\cite{Bern:1993mq,Bern:1994fz} as well as \texttt{OpenLoops2}~\cite{Buccioni:2019sur}, by numerically evaluating all helicity configurations of the channels listed in \eqref{eq:channels_all} and \eqref{eq:channels_sameflavor}.
Finally, we found perfect agreement with the analytic two-loop full-color all-plus gluon amplitude of ref.~\cite{Badger:2019djh} and with the numerical benchmarks in the leading-color approximation provided in ref.~\cite{Abreu:2021oya} for all channels in \eqref{eq:channels_all}.

In table \ref{tab:numerical_results} we present benchmark results for 
all the relevant partonic channels. We provide the finite remainder of the squared matrix elements summed over color and helicities, normalized by the corresponding leading order term.
We choose the kinematic configuration
\begin{alignat}{3}
    & s_{12} = 10^6,      \quad && s_{23} = -761244.13, \quad && s_{34} = 865719.14, \notag \\
    & s_{45} = 126204.05, \quad && s_{51} = -29885.560, \quad && \mu^2 = 10^4,
\end{alignat}
and fix $N_c=3$ and $N_f=5$. To evaluate  the pentagon functions we use the $\texttt{PentagonMI}$ package~\cite{Chicherin:2020oor}.
All results in electronic format are available at~\cite{results:url}.
\newline

\begingroup
\begin{table}[t!]
    \renewcommand{\arraystretch}{1.5}
    \centering
    \begin{tabular}{c|cccc}
         & \;$ \overline{\sum} 2 {\rm Re}[{\colorvector{A}^0}^\dagger \colorvector{A}^1_\fin]$ \;& $\overline{\sum} |\colorvector{A}^1_\fin|^2$ \;& $\overline{\sum} 2 {\rm Re}[{\colorvector{A}^0}^\dagger \colorvector{A}^2_\fin]$  \\        
        \hline
        $gg \to ggg$       & -90.64321 & 3348.355 & 2856.837  \\
        \hline
        $q\bar{q} \to ggg$ & -115.3289 & 3939.841 & 3833.951 \\
        $q g \to qgg$      & -74.31499 & 1917.467 & 1195.185 \\ 
        $gg \to q\bar{q}g$ & -72.79952 & 3093.624 & 1503.403 \\
        \hline
        $q\bar q\to \bar q'q' g$  &  -101.6531 & 3271.088 & 2511.430 \\
        $q\bar q' \to \bar q' q g$  & -82.09317 & 4338.144 & -768.0230 \\
        $\bar q' \bar q \to \bar q' \bar q g$  & -47.41403 & 769.2739 & 82.75641 \\
        $g \bar q\to  \bar q' q' \bar q $  & -57.86782 & 1181.730 & 1341.638 \\
        $q\bar q \to \bar q q g$ & -88.39101 & 3926.462 & 379.8467 \\
        $q q \to q q g$  & -45.63443 & 767.2815 & 94.00947 \\
        $qg \to qq\bar q$  & -71.33829 & 1686.104 & 1626.085 \\
    \end{tabular}
    \caption{Benchmark results for the interference of the tree-level with the one- and two-loop finite remainders (first and third columns)
    and for the squared one-loop finite remainder (second column). $\overline{\sum}$ refers to summation over color and helicity states, 
    and normalization over the corresponding leading order term.}
    \label{tab:numerical_results}
\end{table}
\endgroup
To conclude, in this Letter we have presented the calculation of the two-loop corrections to five-parton scattering in massless QCD, retaining full color dependence.
Our calculation leveraged many state-of-the-art techniques in the evaluation of multi-loop scattering amplitudes, including the helicity projector technique in the 't Hooft-Veltman scheme,
finite field and syzygy based reduction algorithms,
and multivariate partial-fraction decomposition.
We considered all relevant partonic channels and derived compact analytic results, that can be easily evaluated numerically for physical scattering kinematics.
The amplitudes presented here constitute the last missing building block to obtain full-color NNLO predictions for three-jet observables at the LHC.
Moreover, they can furnish important information to study multi-Regge kinematics in QCD and to investigate collinear factorization breaking~\cite{Catani:2011st,Dixon:2019lnw}.
\newline

\noindent \textbf{Note added}\\
During the final stages of completion of this project, we have become aware of another
concurrent calculation of the gluonic~\cite{DeLaurentis:2023nss}
and quark~\cite{toappear} processes. 
While the two calculations have been performed in two different
infrared subtraction schemes, we have verified that,
after scheme change, the results agree numerically to high precision.
\newline

\begin{acknowledgments}
\textbf{Acknowledgments}
We thank F.\ Caola for enlightening discussions during the
various stages of this project. 
We are grateful to G.\ De Laurentis, H.\ Ita, M.\ Klinkert and V.\ Sotnikov for sharing with
us benchmark results and for pointing out a mistake in some of the numerical results provided 
in table~\ref{tab:numerical_results} in the first version of this article.
This research was supported in part by the Deutsche Forschungsgemeinschaft (DFG, German Research Foundation) under grant 396021762 - TRR 257 and under Germany’s Excellence Strategy - EXC-2094 - 390783311, by the Royal Society grant URF/R1/191125, 
by the National Science Foundation through Grant 2013859 and by the European Research Council (ERC) under the European Union’s research and innovation programme grant agreements ERC Starting Grant 949279 HighPHun and ERC Starting Grant 804394 hipQCD.
\end{acknowledgments}

\bibliographystyle{bibliostyle}
\bibliography{biblio}

\providecommand{\href}[2]{#2}\begingroup\raggedright\begin{thebibliography}{10}

\bibitem{ATLAS:2011qvj}
{\bf ATLAS} Collaboration, G.~Aad et~al., {\it {Measurement of multi-jet cross
  sections in proton-proton collisions at a 7 TeV center-of-mass energy}},
  {\em Eur. Phys. J. C} {\bf 71} (2011) 1763,
  [\href{http://arxiv.org/abs/1107.2092}{{\tt arXiv:1107.2092}}].

\bibitem{CMS:2013vbb}
{\bf CMS} Collaboration, S.~Chatrchyan et~al., {\it {Measurement of the Ratio
  of the Inclusive 3-Jet Cross Section to the Inclusive 2-Jet Cross Section in
  pp Collisions at $\sqrt{s}$ = 7 TeV and First Determination of the Strong
  Coupling Constant in the TeV Range}},  {\em Eur. Phys. J. C} {\bf 73} (2013),
  no.~10 2604, [\href{http://arxiv.org/abs/1304.7498}{{\tt arXiv:1304.7498}}].

\bibitem{CMS:2014mna}
{\bf CMS} Collaboration, V.~Khachatryan et~al., {\it {Measurement of the
  inclusive 3-jet production differential cross section in
  proton\textendash{}proton collisions at 7 TeV and determination of the strong
  coupling constant in the TeV range}},  {\em Eur. Phys. J. C} {\bf 75} (2015),
  no.~5 186, [\href{http://arxiv.org/abs/1412.1633}{{\tt arXiv:1412.1633}}].

\bibitem{ATLAS:2014qmg}
{\bf ATLAS} Collaboration, G.~Aad et~al., {\it {Measurement of three-jet
  production cross-sections in $pp$ collisions at 7 TeV centre-of-mass energy
  using the ATLAS detector}},  {\em Eur. Phys. J. C} {\bf 75} (2015), no.~5
  228, [\href{http://arxiv.org/abs/1411.1855}{{\tt arXiv:1411.1855}}].

\bibitem{Gehrmann-DeRidder:2019ibf}
A.~Gehrmann-De~Ridder, T.~Gehrmann, E.~W.~N. Glover, A.~Huss, and J.~Pires,
  {\it {Triple Differential Dijet Cross Section at the LHC}},  {\em Phys. Rev.
  Lett.} {\bf 123} (2019), no.~10 102001,
  [\href{http://arxiv.org/abs/1905.09047}{{\tt arXiv:1905.09047}}].

\bibitem{Czakon:2019tmo}
M.~Czakon, A.~van Hameren, A.~Mitov, and R.~Poncelet, {\it {Single-jet
  inclusive rates with exact color at $ \mathcal{O} $ ($ {\alpha}_s^4 $)}},
  {\em JHEP} {\bf 10} (2019) 262, [\href{http://arxiv.org/abs/1907.12911}{{\tt
  arXiv:1907.12911}}].

\bibitem{Czakon:2021mjy}
M.~Czakon, A.~Mitov, and R.~Poncelet, {\it {Next-to-Next-to-Leading Order Study
  of Three-Jet Production at the LHC}},  {\em Phys. Rev. Lett.} {\bf 127}
  (2021), no.~15 152001, [\href{http://arxiv.org/abs/2106.05331}{{\tt
  arXiv:2106.05331}}]. [Erratum: Phys.Rev.Lett. 129, 119901 (2022), Erratum:
  Phys.Rev.Lett. 129, 119901 (2022)].

\bibitem{Chen:2022tpk}
X.~Chen, T.~Gehrmann, E.~W.~N. Glover, A.~Huss, and J.~Mo, {\it {NNLO QCD
  corrections in full colour for jet production observables at the LHC}},  {\em
  JHEP} {\bf 09} (2022) 025, [\href{http://arxiv.org/abs/2204.10173}{{\tt
  arXiv:2204.10173}}].

\bibitem{Alvarez:2023fhi}
M.~Alvarez, J.~Cantero, M.~Czakon, J.~Llorente, A.~Mitov, and R.~Poncelet, {\it
  {NNLO QCD corrections to event shapes at the LHC}},  {\em JHEP} {\bf 03}
  (2023) 129, [\href{http://arxiv.org/abs/2301.01086}{{\tt arXiv:2301.01086}}].

\bibitem{Gribov:2009zz}
V.~N. Gribov, {\em {Strong interactions of hadrons at high emnergies: Gribov
  lectures on Theoretical Physics}}.
\newblock Cambridge University Press, 10, 2012.

\bibitem{Agarwal:2021ais}
N.~Agarwal, L.~Magnea, C.~Signorile-Signorile, and A.~Tripathi, {\it {The
  infrared structure of perturbative gauge theories}},  {\em Phys. Rept.} {\bf
  994} (2023) 1--120, [\href{http://arxiv.org/abs/2112.07099}{{\tt
  arXiv:2112.07099}}].

\bibitem{Tkachov:1981wb}
F.~Tkachov, {\it {A Theorem on Analytical Calculability of Four Loop
  Renormalization Group Functions}},  {\em Phys.Lett.} {\bf B100} (1981)
  65--68.

\bibitem{Chetyrkin:1981qh}
K.~Chetyrkin and F.~Tkachov, {\it {Integration by Parts: The Algorithm to
  Calculate beta Functions in 4 Loops}},  {\em Nucl.Phys.} {\bf B192} (1981)
  159--204.

\bibitem{Laporta:2001dd}
S.~Laporta, {\it {High precision calculation of multiloop Feynman integrals by
  difference equations}},  {\em Int.J.Mod.Phys.} {\bf A15} (2000) 5087--5159,
  [\href{http://arxiv.org/abs/hep-ph/0102033}{{\tt hep-ph/0102033}}].

\bibitem{Kotikov:1990kg}
A.~Kotikov, {\it {Differential equations method: New technique for massive
  Feynman diagrams calculation}},  {\em Phys.Lett.} {\bf B254} (1991) 158--164.

\bibitem{Remiddi:1997ny}
E.~Remiddi, {\it {Differential equations for Feynman graph amplitudes}},  {\em
  Nuovo Cim.} {\bf A110} (1997) 1435--1452,
  [\href{http://arxiv.org/abs/hep-th/9711188}{{\tt hep-th/9711188}}].

\bibitem{Gehrmann:1999as}
T.~Gehrmann and E.~Remiddi, {\it {Differential equations for two loop four
  point functions}},  {\em Nucl.Phys.} {\bf B580} (2000) 485--518,
  [\href{http://arxiv.org/abs/hep-ph/9912329}{{\tt hep-ph/9912329}}].

\bibitem{vonManteuffel:2014ixa}
A.~von Manteuffel and R.~M. Schabinger, {\it {A novel approach to integration
  by parts reduction}},  {\em Phys. Lett.} {\bf B744} (2015) 101--104,
  [\href{http://arxiv.org/abs/1406.4513}{{\tt arXiv:1406.4513}}].

\bibitem{Peraro:2019svx}
T.~Peraro, {\it {FiniteFlow: multivariate functional reconstruction using
  finite fields and dataflow graphs}},
  \href{http://arxiv.org/abs/1905.08019}{{\tt arXiv:1905.08019}}.

\bibitem{Henn:2013pwa}
J.~M. Henn, {\it {Multiloop integrals in dimensional regularization made
  simple}},  {\em Phys.Rev.Lett.} {\bf 110} (2013) 251601,
  [\href{http://arxiv.org/abs/1304.1806}{{\tt arXiv:1304.1806}}].

\bibitem{Remiddi:1999ew}
E.~Remiddi and J.~Vermaseren, {\it {Harmonic polylogarithms}},  {\em
  Int.J.Mod.Phys.} {\bf A15} (2000) 725--754,
  [\href{http://arxiv.org/abs/hep-ph/9905237}{{\tt hep-ph/9905237}}].

\bibitem{Goncharov:1998kja}
A.~B. Goncharov, {\it {Multiple polylogarithms, cyclotomy and modular
  complexes}},  {\em Math. Res. Lett.} {\bf 5} (1998) 497--516,
  [\href{http://arxiv.org/abs/1105.2076}{{\tt arXiv:1105.2076}}].

\bibitem{Goncharov:2001iea}
A.~B. Goncharov, {\it {Multiple polylogarithms and mixed Tate motives}},
  \href{http://arxiv.org/abs/math/0103059}{{\tt math/0103059}}.

\bibitem{Goncharov:2010jf}
A.~B. Goncharov, M.~Spradlin, C.~Vergu, and A.~Volovich, {\it {Classical
  Polylogarithms for Amplitudes and Wilson Loops}},  {\em Phys.Rev.Lett.} {\bf
  105} (2010) 151605, [\href{http://arxiv.org/abs/1006.5703}{{\tt
  arXiv:1006.5703}}].

\bibitem{Chen:1977oja}
K.-T. Chen, {\it {Iterated path integrals}},  {\em Bull.Am.Math.Soc.} {\bf 83}
  (1977) 831--879.

\bibitem{Lee:2021uqq}
R.~N. Lee, A.~von Manteuffel, R.~M. Schabinger, A.~V. Smirnov, V.~A. Smirnov,
  and M.~Steinhauser, {\it {Fermionic corrections to quark and gluon form
  factors in four-loop QCD}},  {\em Phys. Rev. D} {\bf 104} (2021), no.~7
  074008, [\href{http://arxiv.org/abs/2105.11504}{{\tt arXiv:2105.11504}}].

\bibitem{Lee:2022nhh}
R.~N. Lee, A.~von Manteuffel, R.~M. Schabinger, A.~V. Smirnov, V.~A. Smirnov,
  and M.~Steinhauser, {\it {Quark and Gluon Form Factors in Four-Loop QCD}},
  {\em Phys. Rev. Lett.} {\bf 128} (2022), no.~21 212002,
  [\href{http://arxiv.org/abs/2202.04660}{{\tt arXiv:2202.04660}}].

\bibitem{Caola:2020dfu}
F.~Caola, A.~Von~Manteuffel, and L.~Tancredi, {\it {Diphoton Amplitudes in
  Three-Loop Quantum Chromodynamics}},  {\em Phys. Rev. Lett.} {\bf 126}
  (2021), no.~11 112004, [\href{http://arxiv.org/abs/2011.13946}{{\tt
  arXiv:2011.13946}}].

\bibitem{Caola:2021izf}
F.~Caola, A.~Chakraborty, G.~Gambuti, A.~von Manteuffel, and L.~Tancredi, {\it
  {Three-Loop Gluon Scattering in QCD and the Gluon Regge Trajectory}},  {\em
  Phys. Rev. Lett.} {\bf 128} (2022), no.~21 212001,
  [\href{http://arxiv.org/abs/2112.11097}{{\tt arXiv:2112.11097}}].

\bibitem{Caola:2021rqz}
F.~Caola, A.~Chakraborty, G.~Gambuti, A.~von Manteuffel, and L.~Tancredi, {\it
  {Three-loop helicity amplitudes for four-quark scattering in massless QCD}},
  {\em JHEP} {\bf 10} (2021) 206, [\href{http://arxiv.org/abs/2108.00055}{{\tt
  arXiv:2108.00055}}].

\bibitem{Abreu:2018zmy}
S.~Abreu, J.~Dormans, F.~Febres~Cordero, H.~Ita, and B.~Page, {\it {Analytic
  Form of Planar Two-Loop Five-Gluon Scattering Amplitudes in QCD}},  {\em
  Phys. Rev. Lett.} {\bf 122} (2019), no.~8 082002,
  [\href{http://arxiv.org/abs/1812.04586}{{\tt arXiv:1812.04586}}].

\bibitem{Abreu:2019odu}
S.~Abreu, J.~Dormans, F.~Febres~Cordero, H.~Ita, B.~Page, and V.~Sotnikov, {\it
  {Analytic Form of the Planar Two-Loop Five-Parton Scattering Amplitudes in
  QCD}},  {\em JHEP} {\bf 05} (2019) 084,
  [\href{http://arxiv.org/abs/1904.00945}{{\tt arXiv:1904.00945}}].

\bibitem{Badger:2013gxa}
S.~Badger, H.~Frellesvig, and Y.~Zhang, {\it {A Two-Loop Five-Gluon Helicity
  Amplitude in QCD}},  {\em JHEP} {\bf 12} (2013) 045,
  [\href{http://arxiv.org/abs/1310.1051}{{\tt arXiv:1310.1051}}].

\bibitem{Badger:2015lda}
S.~Badger, G.~Mogull, A.~Ochirov, and D.~O'Connell, {\it {A Complete Two-Loop,
  Five-Gluon Helicity Amplitude in Yang-Mills Theory}},  {\em JHEP} {\bf 10}
  (2015) 064, [\href{http://arxiv.org/abs/1507.08797}{{\tt arXiv:1507.08797}}].

\bibitem{Badger:2018enw}
S.~Badger, C.~Br\o{}nnum-Hansen, H.~B. Hartanto, and T.~Peraro, {\it {Analytic
  helicity amplitudes for two-loop five-gluon scattering: the single-minus
  case}},  {\em JHEP} {\bf 01} (2019) 186,
  [\href{http://arxiv.org/abs/1811.11699}{{\tt arXiv:1811.11699}}].

\bibitem{Badger:2019djh}
S.~Badger, D.~Chicherin, T.~Gehrmann, G.~Heinrich, J.~Henn, T.~Peraro,
  P.~Wasser, Y.~Zhang, and S.~Zoia, {\it {Analytic form of the full two-loop
  five-gluon all-plus helicity amplitude}},  {\em Phys. Rev. Lett.} {\bf 123}
  (2019), no.~7 071601, [\href{http://arxiv.org/abs/1905.03733}{{\tt
  arXiv:1905.03733}}].

\bibitem{Abreu:2021oya}
S.~Abreu, F.~Febres~Cordero, H.~Ita, B.~Page, and V.~Sotnikov, {\it
  {Leading-color two-loop QCD corrections for three-jet production at hadron
  colliders}},  {\em JHEP} {\bf 07} (2021) 095,
  [\href{http://arxiv.org/abs/2102.13609}{{\tt arXiv:2102.13609}}].

\bibitem{Agarwal:2021vdh}
B.~Agarwal, F.~Buccioni, A.~von Manteuffel, and L.~Tancredi, {\it {Two-Loop
  Helicity Amplitudes for Diphoton Plus Jet Production in Full Color}},  {\em
  Phys. Rev. Lett.} {\bf 127} (2021), no.~26 262001,
  [\href{http://arxiv.org/abs/2105.04585}{{\tt arXiv:2105.04585}}].

\bibitem{Badger:2021ohm}
S.~Badger, T.~Gehrmann, M.~Marcoli, and R.~Moodie, {\it {Next-to-leading order
  QCD corrections to diphoton-plus-jet production through gluon fusion at the
  LHC}},  {\em Phys. Lett. B} {\bf 824} (2022) 136802,
  [\href{http://arxiv.org/abs/2109.12003}{{\tt arXiv:2109.12003}}].

\bibitem{Badger:2023mgf}
S.~Badger, M.~Czakon, H.~B. Hartanto, R.~Moodie, T.~Peraro, R.~Poncelet, and
  S.~Zoia, {\it {Isolated photon production in association with a jet pair
  through next-to-next-to-leading order in QCD}},  {\em JHEP} {\bf 10} (2023)
  071, [\href{http://arxiv.org/abs/2304.06682}{{\tt arXiv:2304.06682}}].

\bibitem{Abreu:2023bdp}
S.~Abreu, G.~De~Laurentis, H.~Ita, M.~Klinkert, B.~Page, and V.~Sotnikov, {\it
  {Two-Loop QCD Corrections for Three-Photon Production at Hadron Colliders}},
  {\em SciPost Phys.} {\bf 15} (2023) 157,
  [\href{http://arxiv.org/abs/2305.17056}{{\tt arXiv:2305.17056}}].

\bibitem{Chicherin:2020oor}
D.~Chicherin and V.~Sotnikov, {\it {Pentagon Functions for Scattering of Five
  Massless Particles}},  {\em JHEP} {\bf 20} (2020) 167,
  [\href{http://arxiv.org/abs/2009.07803}{{\tt arXiv:2009.07803}}].

\bibitem{tHooft:1972tcz}
G.~'t~Hooft and M.~J.~G. Veltman, {\it {Regularization and Renormalization of
  Gauge Fields}},  {\em Nucl. Phys.} {\bf B44} (1972) 189--213.

\bibitem{Catani:1998bh}
S.~Catani, {\it {The Singular behavior of QCD amplitudes at two loop order}},
  {\em Phys.Lett.} {\bf B427} (1998) 161--171,
  [\href{http://arxiv.org/abs/hep-ph/9802439}{{\tt hep-ph/9802439}}].

\bibitem{Becher:2009cu}
T.~Becher and M.~Neubert, {\it {Infrared singularities of scattering amplitudes
  in perturbative QCD}},  {\em Phys. Rev. Lett.} {\bf 102} (2009) 162001,
  [\href{http://arxiv.org/abs/0901.0722}{{\tt arXiv:0901.0722}}]. [Erratum:
  Phys.Rev.Lett. 111, 199905 (2013)].

\bibitem{Becher:2009qa}
T.~Becher and M.~Neubert, {\it {On the Structure of Infrared Singularities of
  Gauge-Theory Amplitudes}},  {\em JHEP} {\bf 06} (2009) 081,
  [\href{http://arxiv.org/abs/0903.1126}{{\tt arXiv:0903.1126}}]. [Erratum:
  JHEP 11, 024 (2013)].

\bibitem{Nogueira:1991ex}
P.~Nogueira, {\it {Automatic Feynman graph generation}},  {\em J.Comput.Phys.}
  {\bf 105} (1993) 279--289.

\bibitem{Peraro:2019cjj}
T.~Peraro and L.~Tancredi, {\it {Physical projectors for multi-leg helicity
  amplitudes}},  {\em JHEP} {\bf 07} (2019) 114,
  [\href{http://arxiv.org/abs/1906.03298}{{\tt arXiv:1906.03298}}].

\bibitem{Peraro:2020sfm}
T.~Peraro and L.~Tancredi, {\it {Tensor decomposition for bosonic and fermionic
  scattering amplitudes}},  {\em Phys. Rev. D} {\bf 103} (2021), no.~5 054042,
  [\href{http://arxiv.org/abs/2012.00820}{{\tt arXiv:2012.00820}}].

\bibitem{Vermaseren:2000nd}
J.~Vermaseren, {\it {New features of FORM}},
  \href{http://arxiv.org/abs/math-ph/0010025}{{\tt math-ph/0010025}}.

\bibitem{Ruijl:2017dtg}
B.~Ruijl, T.~Ueda, and J.~Vermaseren, {\it {FORM version 4.2}},
  \href{http://arxiv.org/abs/1707.06453}{{\tt arXiv:1707.06453}}.

\bibitem{Papadopoulos:2015jft}
C.~G. Papadopoulos, D.~Tommasini, and C.~Wever, {\it {The Pentabox Master
  Integrals with the Simplified Differential Equations approach}},  {\em JHEP}
  {\bf 04} (2016) 078, [\href{http://arxiv.org/abs/1511.09404}{{\tt
  arXiv:1511.09404}}].

\bibitem{Gehrmann:2018yef}
T.~Gehrmann, J.~Henn, and N.~Lo~Presti, {\it {Pentagon functions for massless
  planar scattering amplitudes}},  {\em JHEP} {\bf 10} (2018) 103,
  [\href{http://arxiv.org/abs/1807.09812}{{\tt arXiv:1807.09812}}].

\bibitem{Chicherin:2018mue}
D.~Chicherin, T.~Gehrmann, J.~Henn, N.~Lo~Presti, V.~Mitev, and P.~Wasser, {\it
  {Analytic result for the nonplanar hexa-box integrals}},  {\em JHEP} {\bf 03}
  (2019) 042, [\href{http://arxiv.org/abs/1809.06240}{{\tt arXiv:1809.06240}}].

\bibitem{Studerus:2009ye}
C.~Studerus, {\it {Reduze-Feynman Integral Reduction in C++}},  {\em
  Comput.Phys.Commun.} {\bf 181} (2010) 1293--1300,
  [\href{http://arxiv.org/abs/0912.2546}{{\tt arXiv:0912.2546}}].

\bibitem{vonManteuffel:2012np}
A.~von Manteuffel and C.~Studerus, {\it {Reduze 2 - Distributed Feynman
  Integral Reduction}},  \href{http://arxiv.org/abs/1201.4330}{{\tt
  arXiv:1201.4330}}.

\bibitem{Maierhofer:2017gsa}
P.~Maierh\"ofer, J.~Usovitsch, and P.~Uwer, {\it {Kira\textemdash{}A Feynman
  integral reduction program}},  {\em Comput. Phys. Commun.} {\bf 230} (2018)
  99--112, [\href{http://arxiv.org/abs/1705.05610}{{\tt arXiv:1705.05610}}].

\bibitem{Klappert:2020nbg}
J.~Klappert, F.~Lange, P.~Maierh\"ofer, and J.~Usovitsch, {\it {Integral
  reduction with Kira 2.0 and finite field methods}},  {\em Comput. Phys.
  Commun.} {\bf 266} (2021) 108024,
  [\href{http://arxiv.org/abs/2008.06494}{{\tt arXiv:2008.06494}}].

\bibitem{vonManteuffel:2016xki}
A.~von Manteuffel and R.~M. Schabinger, {\it {Quark and gluon form factors to
  four-loop order in QCD: the $N_f^3$ contributions}},  {\em Phys. Rev.} {\bf
  D95} (2017), no.~3 034030, [\href{http://arxiv.org/abs/1611.00795}{{\tt
  arXiv:1611.00795}}].

\bibitem{Peraro:2016wsq}
T.~Peraro, {\it {Scattering amplitudes over finite fields and multivariate
  functional reconstruction}},  {\em JHEP} {\bf 12} (2016) 030,
  [\href{http://arxiv.org/abs/1608.01902}{{\tt arXiv:1608.01902}}].

\bibitem{Gluza:2010ws}
J.~Gluza, K.~Kajda, and D.~A. Kosower, {\it {Towards a Basis for Planar
  Two-Loop Integrals}},  {\em Phys. Rev. D} {\bf 83} (2011) 045012,
  [\href{http://arxiv.org/abs/1009.0472}{{\tt arXiv:1009.0472}}].

\bibitem{Schabinger:2011dz}
R.~M. Schabinger, {\it {A New Algorithm For The Generation Of
  Unitarity-Compatible Integration By Parts Relations}},  {\em JHEP} {\bf 01}
  (2012) 077, [\href{http://arxiv.org/abs/1111.4220}{{\tt arXiv:1111.4220}}].

\bibitem{Ita:2015tya}
H.~Ita, {\it {Two-loop Integrand Decomposition into Master Integrals and
  Surface Terms}},  {\em Phys. Rev.} {\bf D94} (2016), no.~11 116015,
  [\href{http://arxiv.org/abs/1510.05626}{{\tt arXiv:1510.05626}}].

\bibitem{Larsen:2015ped}
K.~J. Larsen and Y.~Zhang, {\it {Integration-by-parts reductions from unitarity
  cuts and algebraic geometry}},  {\em Phys. Rev.} {\bf D93} (2016), no.~4
  041701, [\href{http://arxiv.org/abs/1511.01071}{{\tt arXiv:1511.01071}}].

\bibitem{Boehm:2017wjc}
J.~B\"ohm, A.~Georgoudis, K.~J. Larsen, M.~Schulze, and Y.~Zhang, {\it
  {Complete sets of logarithmic vector fields for integration-by-parts
  identities of Feynman integrals}},  {\em Phys. Rev. D} {\bf 98} (2018), no.~2
  025023, [\href{http://arxiv.org/abs/1712.09737}{{\tt arXiv:1712.09737}}].

\bibitem{Agarwal:2020dye}
B.~Agarwal, S.~P. Jones, and A.~von Manteuffel, {\it {Two-loop helicity
  amplitudes for $gg \to ZZ$ with full top-quark mass effects}},  {\em JHEP}
  {\bf 05} (2021) 256, [\href{http://arxiv.org/abs/2011.15113}{{\tt
  arXiv:2011.15113}}].

\bibitem{Heller:2021qkz}
M.~Heller and A.~von Manteuffel, {\it {MultivariateApart: Generalized partial
  fractions}},  {\em Comput. Phys. Commun.} {\bf 271} (2022) 108174,
  [\href{http://arxiv.org/abs/2101.08283}{{\tt arXiv:2101.08283}}].

\bibitem{Caola:2020xup}
F.~Caola, K.~Melnikov, D.~Napoletano, and L.~Tancredi, {\it {Noncancellation of
  infrared singularities in collisions of massive quarks}},  {\em Phys. Rev. D}
  {\bf 103} (2021), no.~5 054013, [\href{http://arxiv.org/abs/2011.04701}{{\tt
  arXiv:2011.04701}}].

\bibitem{vonManteuffel:2020vjv}
A.~von Manteuffel, E.~Panzer, and R.~M. Schabinger, {\it {Cusp and collinear
  anomalous dimensions in four-loop QCD from form factors}},  {\em Phys. Rev.
  Lett.} {\bf 124} (2020), no.~16 162001,
  [\href{http://arxiv.org/abs/2002.04617}{{\tt arXiv:2002.04617}}].

\bibitem{Gehrmann:2023jyv}
T.~Gehrmann, P.~Jakub\v{c}\'\i{}k, C.~C. Mella, N.~Syrrakos, and L.~Tancredi,
  {\it {Planar three-loop QCD helicity amplitudes for $V$+jet production at
  hadron colliders}},  \href{http://arxiv.org/abs/2307.15405}{{\tt
  arXiv:2307.15405}}.

\bibitem{Boehm:2020ijp}
J.~Boehm, M.~Wittmann, Z.~Wu, Y.~Xu, and Y.~Zhang, {\it {IBP reduction
  coefficients made simple}},  {\em JHEP} {\bf 12} (2020) 054,
  [\href{http://arxiv.org/abs/2008.13194}{{\tt arXiv:2008.13194}}].

\bibitem{Agarwal:2021grm}
B.~Agarwal, F.~Buccioni, A.~von Manteuffel, and L.~Tancredi, {\it {Two-loop
  leading colour QCD corrections to $q \bar{q} \to \gamma \gamma g$ and $q g
  \to \gamma \gamma q$}},  {\em JHEP} {\bf 04} (2021) 201,
  [\href{http://arxiv.org/abs/2102.01820}{{\tt arXiv:2102.01820}}].

\bibitem{DGPS}
W.~Decker, G.-M. Greuel, G.~Pfister, and H.~Sch\"onemann, ``{\sc Singular}
  {4-2-0} --- {A} computer algebra system for polynomial computations.''
  \url{http://www.singular.uni-kl.de}, 2020.

\bibitem{Edison:2011ta}
A.~C. Edison and S.~G. Naculich, {\it {SU(N) group-theory constraints on
  color-ordered five-point amplitudes at all loop orders}},  {\em Nucl. Phys.
  B} {\bf 858} (2012) 488--501, [\href{http://arxiv.org/abs/1111.3821}{{\tt
  arXiv:1111.3821}}].

\bibitem{Bern:1993mq}
Z.~Bern, L.~J. Dixon, and D.~A. Kosower, {\it {One loop corrections to five
  gluon amplitudes}},  {\em Phys. Rev. Lett.} {\bf 70} (1993) 2677--2680,
  [\href{http://arxiv.org/abs/hep-ph/9302280}{{\tt hep-ph/9302280}}].

\bibitem{Bern:1994fz}
Z.~Bern, L.~J. Dixon, and D.~A. Kosower, {\it {One loop corrections to two
  quark three gluon amplitudes}},  {\em Nucl. Phys. B} {\bf 437} (1995)
  259--304, [\href{http://arxiv.org/abs/hep-ph/9409393}{{\tt hep-ph/9409393}}].

\bibitem{Buccioni:2019sur}
F.~Buccioni, J.-N. Lang, J.~M. Lindert, P.~Maierh\"ofer, S.~Pozzorini,
  H.~Zhang, and M.~F. Zoller, {\it {OpenLoops 2}},  {\em Eur. Phys. J. C} {\bf
  79} (2019), no.~10 866, [\href{http://arxiv.org/abs/1907.13071}{{\tt
  arXiv:1907.13071}}].

\bibitem{results:url}
\url{https://zenodo.org/records/10227683}.

\bibitem{Catani:2011st}
S.~Catani, D.~de~Florian, and G.~Rodrigo, {\it {Space-like (versus time-like)
  collinear limits in QCD: Is factorization violated?}},  {\em JHEP} {\bf 07}
  (2012) 026, [\href{http://arxiv.org/abs/1112.4405}{{\tt arXiv:1112.4405}}].

\bibitem{Dixon:2019lnw}
L.~J. Dixon, E.~Herrmann, K.~Yan, and H.~X. Zhu, {\it {Soft gluon emission at
  two loops in full color}},  {\em JHEP} {\bf 05} (2020) 135,
  [\href{http://arxiv.org/abs/1912.09370}{{\tt arXiv:1912.09370}}].

\bibitem{DeLaurentis:2023nss}
G.~De~Laurentis, H.~Ita, M.~Klinkert, and V.~Sotnikov, {\it {Double-Virtual
  NNLO QCD Corrections for Five-Parton Scattering: The Gluon Channel}},
  \href{http://arxiv.org/abs/2311.10086}{{\tt arXiv:2311.10086}}.

\bibitem{toappear}
G.~De~Laurentis, H.~Ita, M.~Klinkert, and V.~Sotnikov, {\it {To appear}}, .

\end{thebibliography}\endgroup

\newpage
\onecolumngrid
\appendix

\section*{SUPPLEMENTAL MATERIAL}

\subsection*{Integral families}

We provide here the definition our two integrals families. 
We recall that for each family ${\rm fam} = \{A,B\}$ we define the integrals as
\begin{equation}
\mathcal{I}^{\rm fam}_{n_1,...,n_{11}} = 
e^{2 \epsilon \gamma_E }\, \int \prod_{i=1}^2 \left(\frac{d^d k_i}{i \pi^{d/2}}\right) \frac{1}{D_1^{n_1} ... D_{11}^{n_{11}} }\,,
\end{equation}
where $d=4-2\epsilon$ is the space-time dimension, $\gamma_E \sim 0.5772$ is the Euler-Mascheroni constant and
the $D_i$ are the propagators. 
For the two families, the propagators read (see $e.g.$ refs.~\cite{Gehrmann:2018yef,Agarwal:2021grm,Agarwal:2021vdh})
\begin{center}
\begin{tabular}{| c | l | l|}
\hline
Prop.\ den.\ & Family A & Family B  \\
\hline
\hline
$D_1$ & $ k_1^2$ &                                        $k_1^2$    \\
$D_2$ & $(k_1+p_1)^2$ &                              $(k_1-p_1)^2$   \\
$D_3$ & $(k_1+p_1+p_2)^2$ &                      $(k_1-p_1-p_2)^2$   \\
$D_4$ & $(k_1+p_1+p_2+p_3)^2$ &              $(k_1-p_1-p_2-p_3)^2$   \\
$D_5$ & $k_2^2 $ &                                        $k_2^2 $   \\
$D_6$ & $(k_2+p_1+p_2+p_3)^2$ &               $(k_2-p_1-p_2-p_3-p_4)^2$   \\
$D_7$ & $(k_2+p_1+p_2+p_3+p_4)^2$ &       $(k_1-k_2)^2$   \\
$D_8$ & $(k_1 - k_2)^2$ &                               $(k_1-k_2+p_4)^2$  \\
$D_9$ & $(k_1+p_1+p_2+p_3+p_4)^2$ &       $(k_2-p_1)^2$   \\
$D_{10}$ & $(k_2 + p_1 )^2$ &                        $(k_2-p_1-p_2)^2$   \\
$D_{11}$ & $(k_2 +p_1 +p_2)^2$ &                 $(k_2-p_1-p_2-p_3)^2$   \\
\hline 
\end{tabular}
\end{center}
where we suppressed the $+i\varepsilon$ from Feynman's prescription everywhere.
All required crossed families are obtained from the two reference families above by permutations of the external momenta.

\makeatletter
\renewcommand\@biblabel[1]{[#1S]}
\makeatother

\end{document}